# Electric field modulation on negative Poisson's ratio of Two-dimensional arsenic and antimony monolayer by first-principles calculation

*Ximing Rong\*, Xinbo Chen, Weida Chen, Xiuwen Zhang\**

Dr. Ximing Rong, Xinbo Chen, Weida Chen, Prof. Xiuwen Zhang
Shenzhen Key Laboratory of Flexible Memory Materials and Devices, Center for Stretchable Electronics and NanoSensor,
College of Physics and Optoelectronic Engineering, Shenzhen University,
Shenzhen 518060, P. R. China

Dr. Ximing Rong
Key Laboratory of Optoelectronic Devices and Systems of Ministry of Education and Guangdong Province,
College of Optoelectronic Engineering, Shenzhen University,
Shenzhen 518060, P. R. China

\* Corresponding authors. E-mail: xmrong@szu.edu.cn (Ximing Rong), xiuwenzhang@szu.edu.cn (Xiuwen Zhang)

**Keywords:** negative Poisson's ratio, electric field modulation, group V monolayers, first-principles calculation

**Abstract**

Negative Poisson's ratio in monolayer black phosphorus (MBP) has recently attracted lots of attention. Here, we have introduced the auxetic effect in MBP-analog Sb for the first time, and achieve NPR modulations on monolayer As and Sb via first-principles calculation. Comparing with As, the monolayer Sb is phonon unstable. By applying an uniaxial strain along each directions, we discovered a zigzag-vertical reversibility on out-of-plane NPR, and the NPR values for monolayer As and Sb are simulated to be -0.125/-0.172 and -0.036/-0.063 by applying strain along zigzag/vertical directions. The NPR values could be significantly manipulated by applying vertical electrical fields, as increased up to 70.3% for monolayer As and decreased up to 55.6% for monolayer Sb. Such intrinsic NPR and electric field modulation could make those monolayers potential applications in auxetic optoelectronic devices, electrodes and sensors, leading to novel multi-functionalities.

**Introduction**

Comparing with materials with positive Poisson's ratio (PPR), negative Poisson's ratio (NPR) materials are rare in category and distinctive in various properties. NPR stands an auxetic nature from a group of materials, whose lateral directions expand when stretched and shrink when compressed [1]. Such materials and their auxetic effect enhanced shear modulus [2], indentation resistance [3], fracture toughness [4], and accompanied advancement make them widely used in fasteners [5], sensors [6], bulletproof armors and vests enhancements [7], biomedicine [8], tougher composites [9], tissue engineering [10], sports equipment [11], and many other potential applications [12-16].

The reports for NPR materials are originally from bulk-scale materials, and most studies are focus on artificial auxetic materials and engineered microstructures [1,12,13,15,17-22]. Structural engineering brings NPR by some special re-entrant structures, crystal grain or crystal microstructures which forms two coupled orthogonal hinges [1,17-20,23]. However, auxetic effect observed in crystalline $SiO_2$ [24,25], α-$TeO_2$ [26] and other cubic materials [27,28] suggests NPR an intrinsic nature for some crystal structures. Recently, auxetic effect has been investigated and predicted in several two-dimensional (2D) materials, making NPR materials hot prospect applications in low dimensional electromechanical devices and novel multi-functionalities. Up to now, out-of-plane NPR has been discovered in black phosphorus (BP) [29-31], GeS [32], SnSe [33] and monolayer As [34], while in-plane NPR has also been predicted in borophene [35-37], penta-graphene [38,39], penta-$B_2N_4$ [39], penta-$B_4N_2$ [39], $Be_5C_2$ [40], $h\alpha$-silica [41], δ-phosphorene [42] and a class of 1T-type monolayer crystals of group 6-7 transition-metal dichalcogenides (TMD) $MX_2$ (M = Mo, W, Tc, Re; X = S, Se, Te) [43]. Similar to bulk auxetic materials, auxetic effect in 2D NPR materials are basically resulted from their puckered or buckled configuration [29], but the study for TMD auxetic materials shows that the occurrence of auxetic behavior could also contributed by the strong coupling between the chalcogen $p$ orbitals and the intermetal $t_{2g}$-bonding orbitals within the basic triangular pyramid structure unit [43].

For electromechanical application, low dimensional auxetic materials with remarkable mechanical, electronic and optoelectronic properties are imperative, but NPR values for most investigated 2D NPR semiconductor materials are relatively small and unable to be manipulated, hindering their further applications. Therefore, NPR modulation is desirable. Meanwhile, structural and electronic variation during NPR modulation is worth investigating to characterize the mechanical and electronic properties during electromechanical application. In this study, we focus on the NPR modulations for group V monolayer As and Sb using electric field ($E_f$). This study is based on the following facts. First, NPR has been observed in MBP [29] and monolayer As [34], which is supposed to be intrinsic nature and originated from its puckered structure, leading to substantial anisotropy in the mechanical behavior and properties. Furthermore, Sb shared the same column of the periodic table with MBP and monolayer As (as shown in Figure 1), which could probably make it potential auxetic material with MBP-analog structure. Finally, electric and optical properties can be modified by strain inducing [44-46] and electrical field modulation [47], which corresponds to a combined effect among strain, crystal structure, band gap transition, electrical field, and probably NPR.

In this paper, we report for the first time the auxetic effect in monolayer Sb, and achive the manipulations of NPR values of monolayer As and Sb using quantum mechanical first-principles calculations by electric field. Electric field modulations on those materials present obvious band gap variation and structural deformation, showing a significant sensitivity on the electrical properties of external perturbations and modifying probabilities on the mechanical loads. Combined with other remarkable electronic and optoelectronic properties, such intrinsic NPR and electric field modulation make those NPR materials potential applications in auxetic optoelectronic devices, electrodes and sensors.

**Computational Methods**

In this work, all first-principles calculations were performed based on the density functional theory (DFT), as implemented in the Vienna *ab initio* simulation package (VASP) code [48,49]. The projector augmented wave (PAW) potential method [50] was used to describe the interactions between ions and electrons, and the exchange–correlation interactions between electrons were treated within the generalized gradient approximation (GGA–PBE) [51]. The plane wave cut-off energy of 500 eV was used for all structural optimizations and electronic calculations. The K-point meshes were generated according to the Monkhorst–Pack scheme [52] for the Brillouin zones, as $\Gamma$-centered with a $k$-grid density of $0.025 \times 2\pi$ Å$^{-1}$ for ionic interactions and electronic SCF interactions. The convergence criterion for the energy in electronic SCF iterations and the force in ionic step iterations were set to $1.0 \times 10^{-8}$ eV and $1.0 \times 10^{-3}$ eV Å$^{-1}$, respectively. An at least 15 Å vacuum slab was included to impose periodic boundary conditions in the perpendicular direction of monolayer, which is enough to eliminate the interaction between monolayer and its neighboring replicas. The phonon spectrum was calculated using the PHONOPY code [53,54] with the finite displacement method, and a $2 \times 2 \times 1$ supercell with a $k$-grid density of $0.01 \times 2\pi$ Å$^{-1}$ was used to ensure the convergence.

For stretching and compressing procedures, we started with the fully relaxed structures, and a uniaxial engineering strain ($\varepsilon$) was applied in either the zigzag, armchair or vertical direction within ±5%. The strain is defined as $\varepsilon = \Delta L/L$, where $\Delta L$ and $L$ represents the change in lattice and the original lattice in corresponding direction, and it is realized by fixing the lattice parameter to a value different from its balanced value during structure relaxation. The positive and negative values of $\varepsilon$ refer to expansion and compression, respectively. With each uniaxial strain applied, the lattice constant in the transverse directions were fully relaxed through energy minimization until the fore in the transverse directions is minimized, and the resulting strain in the transverse directions is extracted from the fully relaxed structure subjected to an applied strain. Noting that the vertical expansion is hardly achieved, the calculations for positive strain in vertical direction is artificial but valuable for investigating the mechanical properties and anisotropy nature of materials. The Poisson's ratio $v_{ab}$ is calculated from the relationship between applied strain ($\varepsilon_a$) and resulting strain ($\varepsilon_b$) by $v_{ab} = -\varepsilon_b/\varepsilon_a$, and is fitted to function $y = -v_1 x + v_2 x^2 + v_3 x^3$, with $v_1$ as the linear Poisson's ratio, as proposed by Ref [29,34].

**Results and Discussion**

Figure 1 shows the top and side views of monolayer As and Sb. Similar to that of MBP, monolayer As and Sb have 4 atoms per primitive cell, and share the same space group with MBP. The lattice parameters of and PBE band gaps are presented in Table 1. For both monolayer As and Sb, the lattices show anisotropy nature and puckered configuration with two nonequivalent As-As (Sb-Sb) bonds and angles, and structure shows a vertical and anisotropic zigzag (*a*) and armchair (*b*) directions with a nonequivalent vertical height ($\Delta c$) along out-of-plain direction. In order to examine the stability conditions of these monolayers, we carried out the phonon dispersion calculation, as illustrated in Figure 2. For monolayer As, as shown in Figure 2a, there is no imaginary frequency along the whole first Brillouin zone, indicating a dynamically stability on monolayer As. However, an obvious imaginary frequency in z-direction acoustic (ZA) was observed at $\Gamma$ point for monolayer Sb (see Figure 2b), indicating that the Sb lattice may exhibit instability in out-of-plane directions. Such situation has also been observed in borophene, which shows superior mechanical flexibility and in-plane NPR but phonon instability [37].

Due to the anisotropic geometric structure, the structural variation and mechanical properties of monolayer As and Sb are expected to be orientation-dependent. The total energy variations as a function of $\varepsilon$ along zigzag ($\varepsilon_x$), armchair ($\varepsilon_y$) and vertical directions ($\varepsilon_z$) for monolayer As and Sb are presented in Fig. 3a and 3b. The energy change for As and Sb within ±5% shows quadric dependence for all strain directions, indicating that the structures have been restrained without any structural collapsibility or phase transition. For different strain directions, out-of-plane strain ($\varepsilon_z$) has larger energy change comparing with both in-plane directions, suggesting that the vertical structural change is harder to achieve. This is understandable because the interlayer bond needs to overcome the structure deformation in both the vertical and in-plain directions with a vertical stress, sustaining a larger resistance from both zigzag and armchair directions. Such situation doesn't exist in in-plain strain because the deformation in out-of-plain direction can be freely varied due to the large freedom in vacuum. For in-plane strain, energy variations via $\varepsilon_x$ is larger than that of $\varepsilon_y$, indicating in-plane anisotropy and a larger elastic constant in zigzag than armchair, which is in consistent with MBP [58]. The band gap variation of monolayer As for different strain directions are shown in Fig. 3c, showing idea band gap regardless of strain directions. With considerable band gap modulations (0.1 eV ~ 1.2 eV) in all three directions, monolayer As can be well suitable for 2D flexible optoelectronic applications. Comparing with As, monolayer Sb has relatively small modulation on band gap within ±5% in in-plane directions (0.07 eV ~ 0.3 eV), and positive strain in out-of-plane directions could cause the structure metallization at ~3%. Combined with phonon instability nature, monolayer Sb could have limit multi-functional applications as electromechanical devices.

Figure 4 shows the structural response for uniaxial strain along different directions for monolayer As and Sb. Commonly, the expansion in one direction could cause shrink in lateral directions, as $\varepsilon_y$ performed (see Figure 4b and 4e) with PPR, but auxetic effect was observed for monolayer As and Sb when strain was applied in zigzag and vertical directions, as $\varepsilon_x$ caused vertical auxeticity (see blue curves in Figure 4a and 4d) and $\varepsilon_z$ caused zigzag auxeticity (see red curves in Figure 4c and 4f), showing a mechanical reversibility. For strain along zigzag direction of monolayer As, the vertical auxeticity increased linearly as $\varepsilon_x$ increased from -5% to 5%, but for strain along vertical direction, the zigzag auxeticity increased but tended to be converged over ~0.04. This is mainly caused by auxetic infinities in vertical direction and auxetic limitation in in-plain direction. Nevertheless, the auxetic effect is significant within ±5%, and is encouraging for multi-functionals. For monolayer Sb, the auxeticities both got converged at around free-standing structures ($\varepsilon$~0), indicating that the auxetic effect is limited comparing with that of monolayer As. By fitting the curve of response strain as a function of applied strain, the NPR values for monolayer As/Sb are calculated to be $v_{xz}$ = -0.125/-0.036, $v_{zx}$ = -0.172/-0.063. Note that the NPR value for As in Ref [34] by SIESTA method is $v_{xz}$ = -0.093, our result is comparable and our method is proved to be credit. Meanwhile, we enrich the NPR family by introducing auxetic effect in monolayer Sb (with $v_{xz}$ = -0.036) for the first time, which is comparable with the NPR value of MBP ($v_{xz}$ = -0.027) [29].

Next, we investigate the electric field ($E_f$) influence on the response of lattice parameters and band gaps for monolayer As and Sb, as shown in Table 1. For in-plane lattice, monolayer As and Sb show expansion in $a$ and $\Delta c$, but shrink in $b$ as $E_f$ increase from 0 to 6 V/nm, suggest that electric field could also cause auxetic effect in zigzag and vertical directions, which are in agreement with the auxetic directions by strain caused NPR. The band gaps for monolayer As decrease as the function of $E_f$, especially with dramatically decrease when $E_f$ has been loaded to 6 V/nm. Interestingly, the band gaps for monolayer Sb increase with $E_f$ increase from 0 to 4 V/nm. But when $E_f$ has been loaded to 6 V/nm, the structure reaches electric breakdown as the gap becomes 0. Meanwhile, the lattice in zigzag direction increased from 4.354 Å to 4.480 Å, while the lattice in armchair direction decreased from 4.892 Å to 4.577 Å, making the structures more 'zig-zig'. Due to the electric break down, we ignored the NPR modulation for monolayer Sb on such $E_f$ load in the following investigation.

Figure 5 shows the response of monolayer As and Sb under different $E_f$ load and uniaxial stress along $x$ (Figure 5a for As and Figure 5c for Sb) and $z$ (Figure 5b for As and Figure 5d for Sb) directions. For monolayer As, the $E_f$ load enhanced the auxetic effect gradually, as the slopes of $\varepsilon_x$-$\varepsilon_z$ and $\varepsilon_z$-$\varepsilon_x$ function are slightly increased when $E_f$ increased from 0 to 4 V/nm and dramatically increased when $E_f$ increased to 6 V/nm (see Figure 5a and 5b), indicating that NPR values can be modified using electric field modulation. However, for monolayer Sb, the slopes become gradually decreased by increasing $E_f$ load from 0 to 4 V/nm, performing different modulation characters from As. Different from the monotonic increase of As, monolayer Sb present a rather flat variation in stretching procedure (ie. positive applied strain), causing a relatively small NPR comparing with the sharp increase of monolayer As, and showing a spatial stability in transverse direction even if the structure has been starched to a considerable strain range. By fitting the strain response relations to function $y = -v_1 x + v_2 x^2 + v_3 x^3$, we simulated the NPR values ($v_1$) by different $E_f$ load, as presented in Table 2. By applying electric field to monolayer As, the NPR values for vertical auxetic could be increased by up to 32% (by applying strain along zigzag direction) and 70.3% (by applying strain along vertical direction) with $E_f$ increased up to 6 V/nm, respectively. For monolayer Sb, the maximum decrease ratios are 27.8% and 55.6%. This is the first time that NPR modulation has been achieved, and such large modulations on NPR are encouraging because this could probably enhance the performance of materials in tissue engineering, auxetic optoelectronic devices, sensors, structural engineering materials and multi-functional applications. Thus, it is reasonable to predict as follows: (1) NPR could be manipulated for most semi-conductive monolayers. (2) The ratio of NPR increase or decrease could be further enlarged by applying larger $E_f$ load as long as the monolayer has not reached the electric breakdown. (3) NPR could probably make inversion if NPR decreased by $E_f$ load, meaning that some PPR monolayer could have opportunity to have NPR character by electric field modulations.

**Conclusion**

In summery, we have introduced a phonon unstable monolayer Sb into the NPR family, and achieve NPR modulations on monolayer As and Sb via first-principles calculation. Monolayer Sb has the same structure as MBP, with an adequate out-of-plane NPR of -0.036. By applying vertical electrical fields, NPR values for zigzag-vertical reversibility increased for monolayer As and decreased for monolayer Sb gradually the electric field increased. Combined with other remarkable electronic and optoelectronic properties, such intrinsic NPR and electric field modulation make those NPR materials potential applications in auxetic optoelectronic devices, electrodes and sensors.

**Conflicts of interest**
There are no conflicts to declare.

**Acknowledgements**
The present work was supported by National Natural Science Foundations of China (Grant No. 11774239, 61827815), Shenzhen Science and Technology Innovation Commission (Grant No. KQTD20170810105439418, JCYJ20170818093035338, JCYJ20170412110137562, ZDSYS201707271554071), Natural Science Foundation of SZU (Grant No. 827-000242), and High-End Researcher Startup Funds of SZU (Grant No. 848-0000040251).

**Figures and tables**

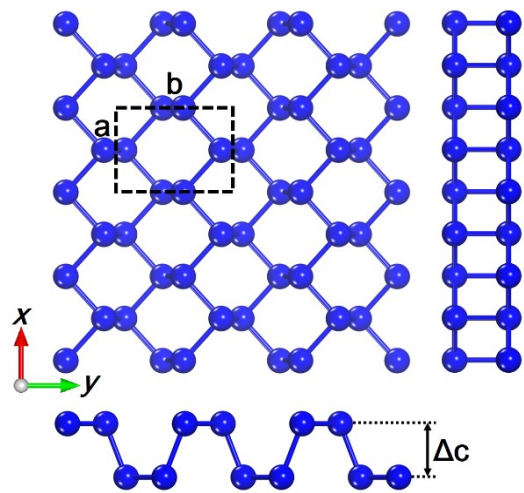

**Figure 1.** Top and side views of As (Sb) monolayer. The unit cell is shown by the dotted rectangle.

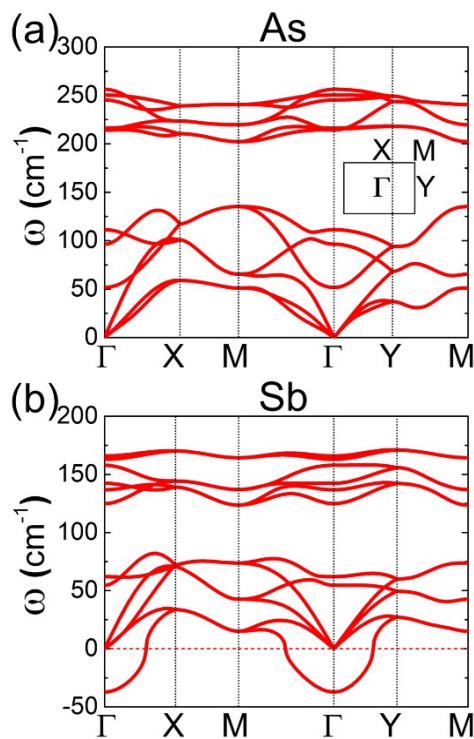

**Figure 2.** Phonon dispersion for monolayer As (a) and Sb (b). Inset in (a) shows the first Brillouin zone.

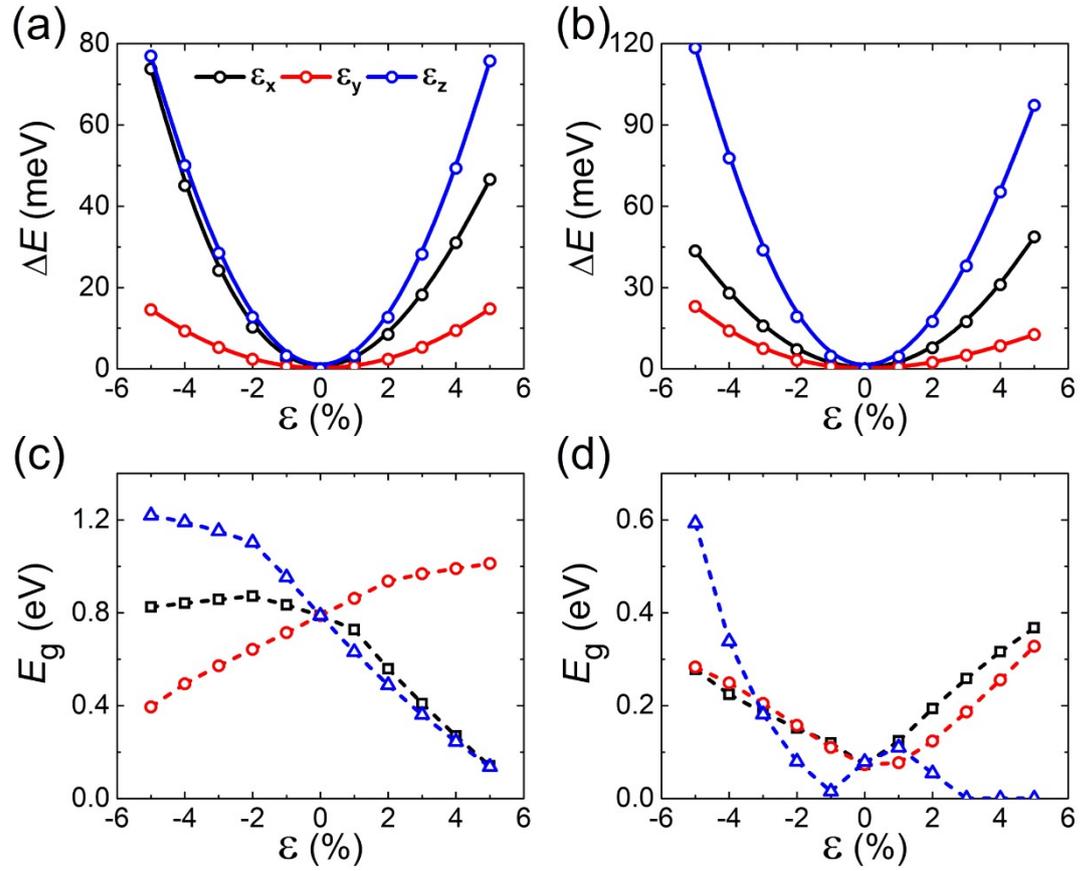

**Figure 3.** The change in the total energy (a/b) band gap (c/d) as a function of applied strain for monolayer As/Sb.

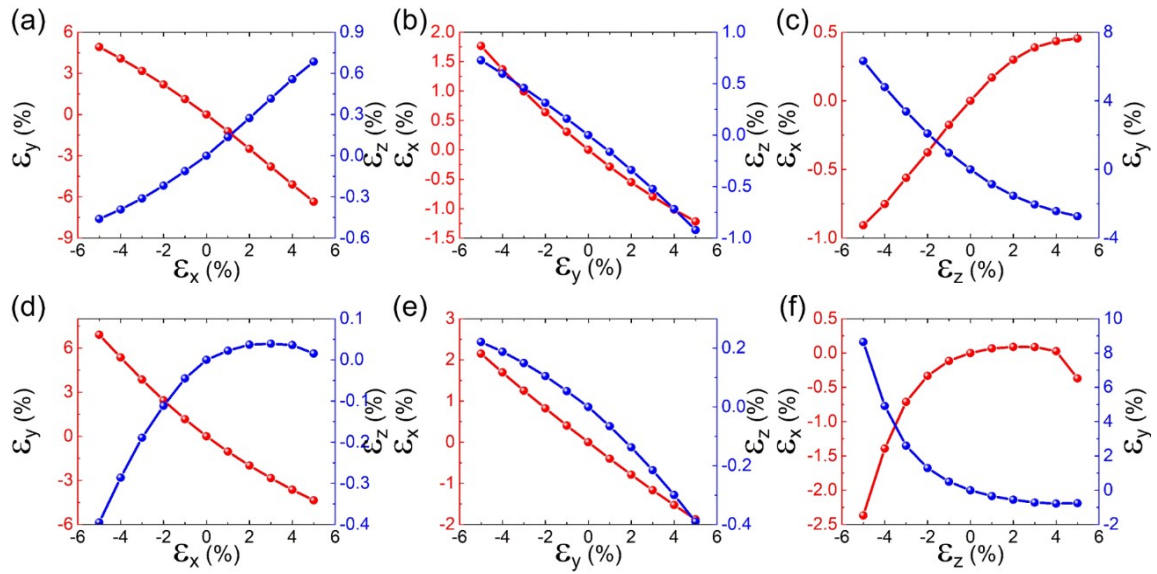

**Figure 4.** Response of monolayer As/Sb under uniaxial stress along x (a/d), y (b/e) and z (c/f)

directions.

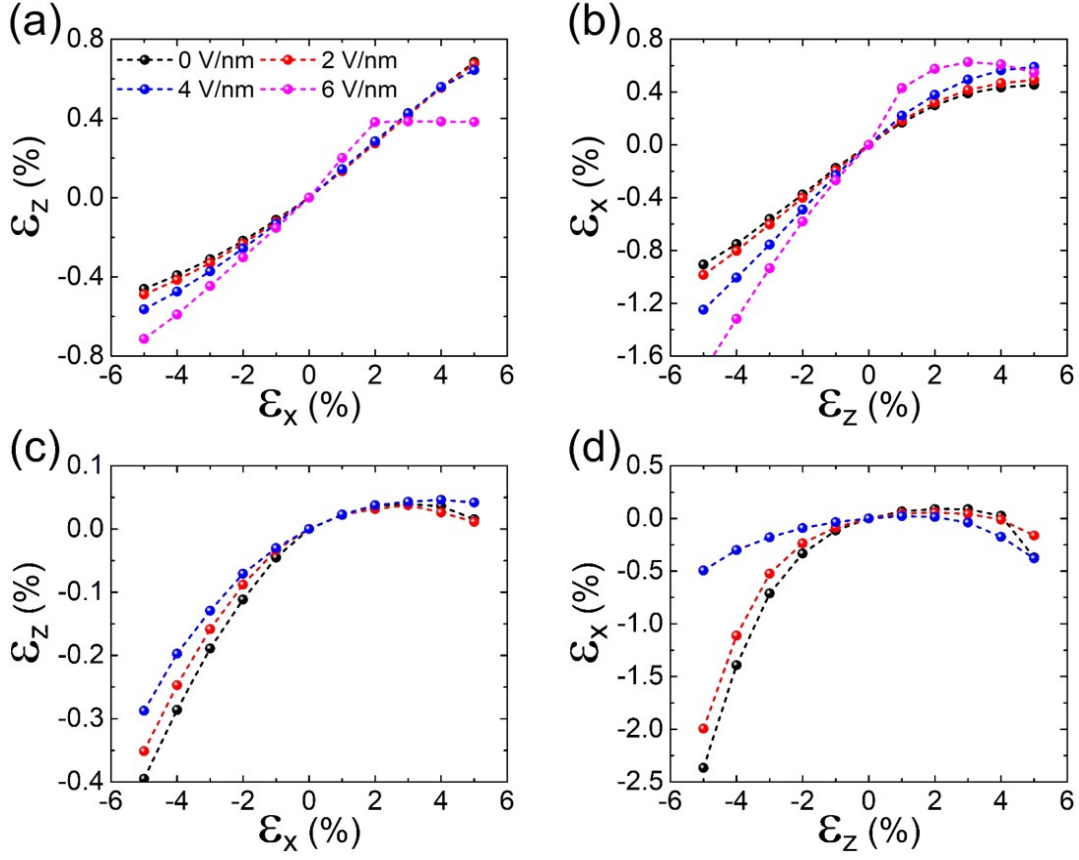

**Figure 5.** Response of monolayer As/Sb under different $E_f$ load and uniaxial stress along x (a/c) and z (b/d) directions.

**Table 1.** Electrical field modulation on the lattice parameters and PBE band gap for monolayer As and Sb.

|      | $E_f$ (V/nm) | a (Å) | b (Å) | $\Delta$c (Å) | $E_g$ (eV) |
|------|--------------|-------|-------|---------------|------------|
| As   | 0            | 3.687 | 4.762 | 2.402         | 0.79       |
|      | 2            | 3.694 | 4.743 | 2.408         | 0.73       |
|      | 4            | 3.720 | 4.680 | 2.428         | 0.55       |
|      | 6            | 3.782 | 4.536 | 2.467         | 0.12       |
| Sb   | 0            | 4.354 | 4.892 | 2.823         | 0.07       |
|      | 2            | 4.371 | 4.857 | 2.832         | 0.08       |
|      | 4            | 4.427 | 4.752 | 2.867         | 0.14       |
|      | 6            | 4.480 | 4.577 | 2.988         | 0          |

**Table 2.** Fitted values and NPR increase (negative for decrease) of $v_{ac}$ and $v_{ca}$ under different $E_f$ load for monolayer As and Sb.

|      | $E_f$ (V/nm) | $v_{ac}$ | $\Delta v_{ac}$(%) | $v_{ca}$ | $\Delta v_{ca}$(%) |
|------|--------------|----------|--------------------|----------|--------------------|
| As   | 0            | -0.125   | 0                  | -0.172   | 0                  |
|      | 2            | -0.129   | 3.2                | -0.184   | 7                  |
|    | 4 | -0.140 | 12    | -0.223 | 30    |
|    | 6 | -0.165 | 32    | -0.293 | 70.3  |
| Sb | 0 | -0.036 | 0     | -0.063 | 0     |
|    | 2 | -0.030 | -16.7 | -0.051 | -19   |
|    | 4 | -0.026 | -27.8 | -0.028 | -55.6 |